\documentclass[mathleft
]{an}
\usepackage{graphicx}
\usepackage{times}
\newcommand{\new}[1]{{ #1}}

\begin{document}

\Pagespan{789}{}
\Yearpublication{2013}%
\Yearsubmission{2013}%
\Month{11}%
\Volume{999}%
\Issue{88}%

\title{A photometric study of the nova-like variable TT\,Arietis with the {\em MOST} satellite}

\author{N. Vogt\inst{1}\fnmsep\thanks{Corresponding author:
  \email{nikolaus.vogt@uv.cl}\newline}
\and  A.-N. Chen\'e\inst{1,2,3}
\and  A.~F~J. Moffat\inst{4}
\and J.~M. Matthews\inst{5}
\and R. Kuschnig\inst{6}
\and D.~B. Guenther\inst{7}
\and J.~F. Rowe\inst{8}
\and S.~M. Rucinski\inst{9}
\and D. Sasselov\inst{10}
\and W.~W. Weiss\inst{6}
}
\titlerunning{A photometric study of TT\,Arietis}
\authorrunning{ N. Vogt et al.}
\institute{ 
Departamento de F\'{\i}sica y Astronom\'{\i}a, Universidad de Valpara\'{\i}so, Av. Gran Breta\~na 1111, Valpara\'{\i}so, Chile
\and 
Departamento de Astronom\'{\i}a, Universidad de Concepci\'on, Casilla 160-C, Chile
\and 
Gemini Observatory, Northern Operations Center, 670 North A'ohoku Place, Hilo, HI 96720, USA
\and 
D\'epartement de physique, Universit\'e de Montr\'eal, C.P. 6128, Succ. Centre-Ville, Montr\'eal, QC, H3C 3J7 \& Centre de Recherche en Astrophysique du Qu\'ebec, Canada
\and 
Dept. of Physics \& Astronomy, Univ. of British Columbia, 6224 Agricultural Rd., Vancouver, BC, V6T 1Z1, Canada
\and 
Institut f\"ur Astronomie, Universit\"at Wien, T\"urkenschanzstrasse 17, A-1180 Vienna, Austria
\and 
Dept. of Astronomy \& Physics, Saint Mary's University, Halifax, NS, B3H 3C3, Canada
\and 
NASA Ames Research Center, Moffett Field, CA 94035
\and 
Dept. of Astronomy \& Astrophysics, Univ. of Toronto, Toronto, ON, M5S 3H4, Canada
\and 
Harvard-Smithsonian Center for Astrophsics, 60 Garden St., Cambridge, MA 102138, USA}

\received{30 Sep 2012}
\accepted{11 Nov 2005}
\publonline{later} 


\abstract{%
Variability on all time scales between seconds and decades is typical for cataclysmic variables (CVs). One of the brightest and best studied CVs is TT\,Ari, a nova-like variable which belongs to the VY Scl subclass, characterized by occasional low states in their light curves. It is also known as a permanent superhumper at high state, revealing ``positive'' (P$_{\rm{S}}$ $>$ P$_{\rm{0}}$) as well as ``negative'' (P$_{\rm{S}}$ $<$ P$_{\rm{0}}$) superhumps, where P$_{\rm{S}}$ is the period of the superhump and P$_{\rm{0}}$ the orbital period. TT\,Ari was observed by the Canadian space telescope {\it MOST} for about 230 hours nearly continuously in 2007, with a time resolution of 48 seconds. Here we analyze these data, obtaining a dominant ``negative'' superhump signal with a period P$_{\rm{S}}$ =  0.1331 days and a mean amplitude of 0.09 mag. Strong flickering with amplitudes up to 0.2 mag and peak-to-peak time scales of 15--20 minutes is superimposed on the periodic variations. We found no indications for significant quasi-periodic oscillations with periods around 15 minutes, reported by other authors. We discuss the known superhump behaviour of TT\,Ari during the last five decades and conclude that our period value is at the upper limit of all hitherto determined ``negative'' superhump periods of TT\,Ari, before and after the {\it MOST} run. 
}
\maketitle

\section{Introduction}
Cataclysmic variables \new{(CVs: Warner \cite{Wa95}, Hellier \cite{He01})} are ideal objects to study the physics of accretion: compared with other accreting objects the dynamical time-scale on which accretion proceeds is relatively short and thus more tractable.  Time-dependent variations of CVs open the possibility to study different mass transfer and accretion conditions within the same binary configuration, a ``natural laboratory'' for accretion disk physics. However, long-term earth-bound observations often suffer from necessary gaps in coverage and, therefore, biases in any period determination. The Microvariability and Oscillations of STars  ({\it MOST}) satellite offers a unique possibility to avoid this problem, due to its uninterrupted observing mode over several weeks. 

One of the most enigmatic characteristics displayed by many CVs is the so-called ``superhump phenomenon'', corresponding to periodic variations with amplitudes of about $\leq$0.2 mag and periods P$_{\rm{S}}$ a few percent different from the orbital binary period P$_{\rm{0}}$. Although this behaviour has been known for several decades, there is still no convincing explanation for it. Superhumps were first detected in SU UMa type dwarf novae (Vogt \cite{Vo74}; \cite{Vo80}), but sometimes they are also observed in nova-like stars (``permanent superhumpers''; Patterson \cite{Pa99}).  While dwarf novae show mostly superhump periods a few percent longer than the orbital period (P$_{\rm{S}}$ $>$ P$_{\rm{0}}$ positive superhumps), in nova-like stars we find positive as well as negative (P$_{\rm{S}}$ $<$ P$_{\rm{0}}$) superhumps, sometimes even in the same star. This is the case for TT\,Ari. 

TT\,Ari was observed by MOST in 2007 for a total of 10 days. Only a brief conference proceedings report (Weingrill et al. \cite{We09}) and a popular article in the German journal ``Sterne und Weltraum'' (Kleinschuster \& Weingrill \cite{Kl11}) have been published so far, based on these observations. The present article re-analyzes these MOST data, with the aim of contributing towards a better understanding of the superhump phenomenon. 

\section{TT\,Ari, a permanent superhumper}

Being a nova-like CV with a high accretion rate, TT\,Ari is one of the brightest CV stars in the sky. It belongs to the particular sub-class of VY Scl-stars, characterized by occasional fadings in brightness by several magnitudes. During these ``low states'' TT\,Ari can be as faint as 16\,th mag, while its normal brightness in the $V$ band is about 10.5\,mag (``high state''). During past decades, these low states repeated every 20 to 25 years and can last between 500 and more than 2000 days, including decline and rise. 

The visual flux of TT\,Ari is dominated by the radiation of the accretion disk which surrounds the white dwarf. \new{From spectroscopic observations of this disk, Wu et al. (\cite{Wu02}) derived a precise orbital period (P$\rm{_0} = 0.13755040\pm1.7\cdot10^{-7}$\,days), a mass ratio of red to white dwarf of $0.20 \pm 0.03$ and an orbital inclination of 29 $\pm$ 6$^\circ$.}

Of special interest is the photometric behaviour of TT\,Ari, rather well documented during the past five decades. The low inclination is compatible with there being no eclipses, humps, ellipsoidal variations or other manifestations varying with orbital phase. However, several authors have reported varying superhump periodicities, in particular negative superhumps between 1961 and 1996, positive superhumps between 1997 and 2004 and again negative superhumps after 2005. A detailed discussion of this behaviour during the last 5 decades is given in section  \ref{Discussion}.


\section{{\it MOST} observations}

{\it MOST} is a Canadian space telescope designed for asteroseismology, but also well-suited to observe any kind of variable stars ($V<12$\,mag) at all time scales between a few minutes and several weeks. The satellite was launched on 2003 June 30 into a polar, Sun-synchronous circular orbit with altitude 850 km and orbital period 101.4 minutes.  It carries a 15-cm Rumak-Maksutov telescope with a single broadband optical filter (350 to 750 nm, centered at 525 nm) attached to a CCD photometer.  Its orbit permits uninterrupted monitoring of targets around the celestial equator ($-19^\circ \leq \delta \leq +36^\circ$). 

TT\,Ari was observed by {\it MOST} for a total of 10 days, between 2007 October 24 and November 3. Individual exposures were 3.03 seconds each with 16 consecutive images stacked on board the satellite, giving a time interval between co-added measurements of 48.5 seconds. Each data point was converted into magnitude scale and the general mean value of all measurements during the 10 days run served as the zero point of this magnitude scale. For details on the observation and data reduction procedure of {\it MOST} data, see Rowe et al. (\cite{Ro06}). 

\section{The superhump ephemeris}

\begin{figure}
\includegraphics[width=8.3cm]{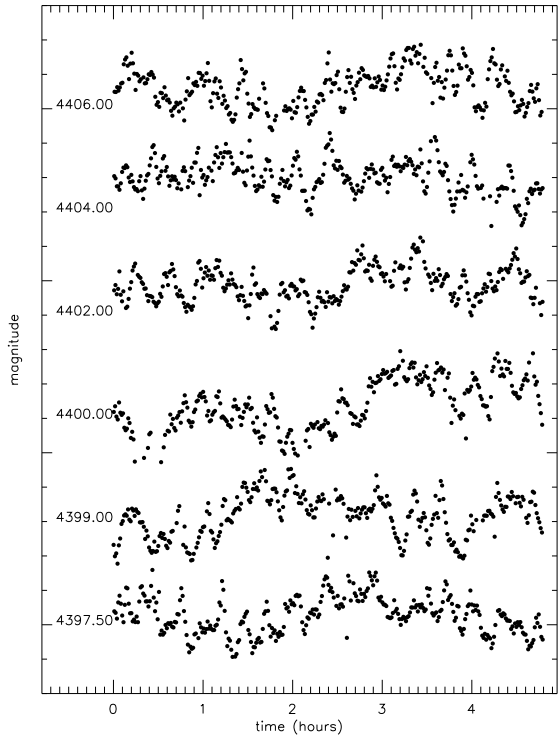}
\caption{A selection of typical sub-sections of the MOST light curve of TT Ari. The magnitude scale is marked in units of 0.1 mag, with arbitrary zero points. The numbers indicate  HJD-2450000  of the first data  point of each light curve. A remarkable flickering activity with typical amplitudes of  0.1 mag and peak to peak time scales of 15--20 minutes is superimposed on the much slower superhump variability. }
\label{lcex}
\end{figure}

The {\it MOST} light curve is dominated by periodic superhump variations, with pronounced maxima coherent over the entire observing run.  The total amplitude of these variations varies between 0.10 and 0.15\,mag. The maxima are normally rather broad while there are often well-defined narrow minima about half-way between subsequent superhump maxima. Figure \ref{lcex} shows some excerpts of the original MOST light curve. A total of 53 maxima and 49 minima are listed in Table \ref{tabsup},  determined by visual inspection of the light curve. The  maxima are given only to three digits, because superimposed flickering does not allow any more accurate determination. Minima are normally narrow and well defined; they are listed to four digits. In both cases the uninterrupted observing mode allowed a unique assignment of cycle count numbers E.

A linear least squares' fit of the moments of maxima reveals
\begin{eqnarray}\label{supmax}
\nonumber\rm{HJD(max)} =  2454397.217(4) + 0.13305(8) \, E\\
\end{eqnarray}

\noindent with a standard deviation of 0.013 days. The corresponding ephemeris for the minima is
\begin{eqnarray}\label{supmin}
\nonumber\rm{HJD(min)} = 2454397.151(3)  +  0.13309(6) \, E\\        
\end{eqnarray}

\noindent with a standard deviation of 0.009\,days. The smaller error of period and standard deviation of the minima is in accordance with their better definition and shorter duration compared to the maxima. There is no significant difference between the periods in (\ref{supmax}) and (\ref{supmin}); the shift in the zero points implies a slight asymmetry of the superhump light curve, with the minima occuring on average at phase 0.58 of the superhump maximum ephemeris (\ref{supmax}). The number of entries in Table \ref{tabsup} is smaller than the total number of cycles covered by {\it MOST}. Even if we take account of two minor gaps in the coverage of about 0.35\,days (= 2.6 cycles of TT\,Ari), only 76\% of the maxima and 71\% of the minima expected from the ephemeris could be identified and listed in Table \ref{tabsup}. This reflects rather strong variability in the light curve shape of individual cycles, while the overall periodicity is rather stable. There is no indication for any significant change of the superhump period within the {\it MOST} observing run. 

Superimposed on these periodic variations there is a pronounced flickering with typical amplitudes of 0.1\,mag and peak-to-peak time scales between 15 and 20 minutes (see Figure \ref{lcex}). Part of this flickering seems to be similar to the quasi-periodic oscillations (QPOs) with periods around 20 minutes (Kraicheva et al. \cite{Kr99}). However, they are not strictly periodic, as shown by the detailed analysis presented in the next section.

\section{Frequency analysis}

\begin{figure}
\includegraphics[width=8cm]{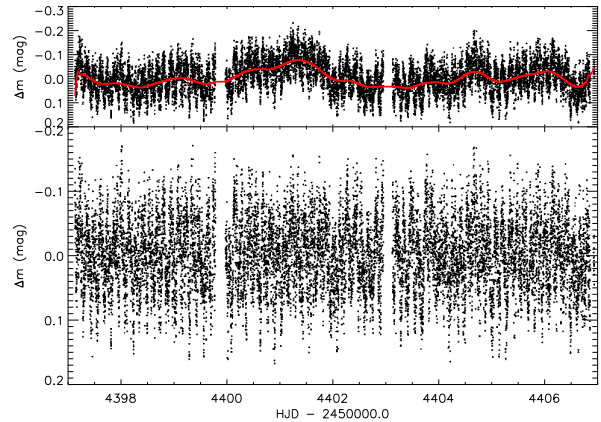}
\caption{{\it Upper panel}:~Raw {\it MOST} light curve of TT\,Arietis. The red central line shows means in time bins of 0.5 days. {\it Lower panel}:~Prewhitened  light curve, using the mean curve showed in the upper panel to subtract low frequency, stochastic-like variations.}
\label{prew}
\end{figure}

\begin{figure}
\includegraphics[width=8cm]{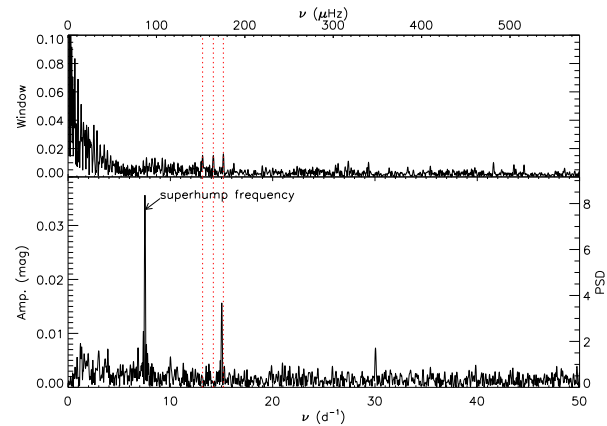}
\caption{{\it Upper panel}:~Window function of the {\it MOST} observations. {\it Lower panel}:~Fourier amplitude spectrum of the prewhitened {\it MOST} light curve in Figure\,\ref{prew}. The 7.513\,d$^{-1}$ ($=86.96\mu$Hz) superhump frequency is marked by an arrow. \new{Harmonics at 15.025\,d$^{-1}$ ($=173.90\mu$Hz) and 30.050\,d$^{-1}$ ($=347.80\mu$Hz) are also visible. Red dotted lines mark the {\it MOST} orbital frequency ($f_{orb}=$ 14.20\,d$^{-1}=164.35\mu$Hz) and its highest sidelobes. The periodogram is marked in both amplitude (mag) and Power Spectrum Density (PSD) units.}}
\label{peri}
\end{figure}

For a detailed analysis of the periodicities at different time scales we have used the Fourier methods as applied to the {\it MOST} data of the WN5-6b star WR\,110 (HD\,165688) by Chen\' e et al. (\cite{Ch11}).  First we determined slow variations, calculating the means in time bins of 0.5 days (Figure \ref{prew}, upper panel). The corresponding residuals (Figure \ref{prew}, lower panel) have been analyzed performing a search for significant signals at frequencies over a range from 2/10\,d$^{-1}$  to 2/1634\,d$^{-1}$ (= 2/50\,s$^{-1}$) with a step of 0.0051 d$^{-1}$, respecting the Nyquist criterion of our run of 10 days duration and about 50 seconds time resolution. \new{This corresponds to a range between $2.3\mu$Hz and $1.4\times10^{-2}\mu$Hz with a step of $5.9\times10^{-2}\mu$Hz. Figure \ref{peri} (lower panel) shows the resulting Fourier amplitude spectrum which is dominated by the superhump signal at $7.513 \pm 0.002 \rm{d}^{-1}$ ($= 86.96 \pm 0.02 \mu$Hz) and its harmonics at $15.025 \pm 0.03 \rm{d}^{-1}$ and $30.050 \pm 0.001 \rm{d}^{-1}$ ($= 173.90 \pm 0.03\mu$Hz and $347.80 \pm 0.01 \mu$Hz, respectively). Note that none of these frequencies fall within or close to the modulation of stray light with the 0.07\,d ($f_{orb}=$ 14.20\,d$^{-1}=164.35\mu$Hz) {\it MOST} orbital period and its harmonics. The closest is the first harmonic of the superhump which is as far as 0.8\,d$^{-1}$, which is not even close to a modulation sidelobe (see the top panel of Figure\,\ref{peri}). The} superhump period $0.133103 \pm 0.000036$\,d is identical with the values derived in (\ref{supmax}) and (\ref{supmin}) in the previous section, considering their errors.

\begin{figure}
\includegraphics[width=8cm]{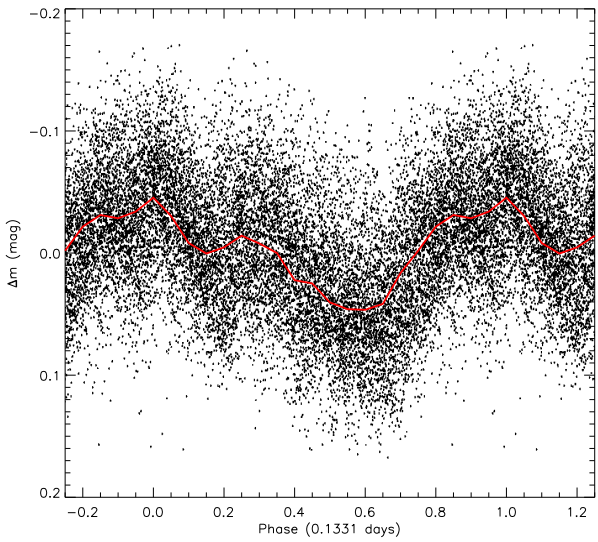}
\caption{Phased light curve from Figure\,\ref{prew} using  P\,=\,0.133050\,days and $t_0$=2454397.2208. The red curve show means in phase bins of 0.05.}
\label{phas}
\end{figure}

Figure \ref{phas} shows the phased light curve using the above period. The smoothed superhump light curve (in phase bins of 0.05) reveals a rather complex variability with a considerable scatter and with mean total amplitude of 0.09\,mag. The minimum occurs around phase 0.6, in accordance with the phase difference between eq. (\ref{supmax}) and (\ref{supmin}). 

\begin{figure}
\includegraphics[width=8cm]{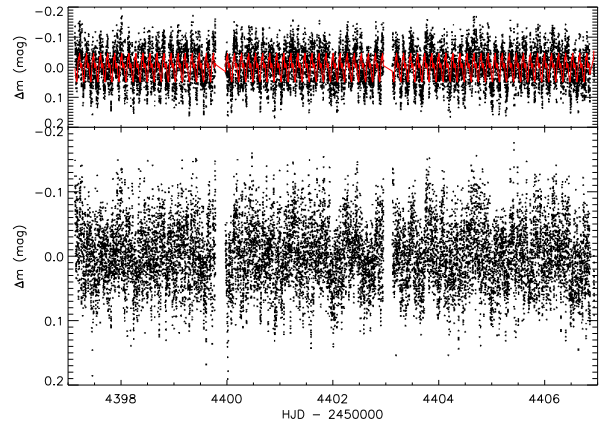}
\caption{{\it Upper panel}:~Prewhitened {\it MOST} light curve from Figure\,\ref{prew} with the superhump variations shown from Figure\,\ref{phas} overplotted. {\it Lower panel}:~Residuals obtained by subtracting the superhump variations from the prewhitened light curve.}
\label{resi}
\end{figure}

\begin{figure}
\includegraphics[width=8cm]{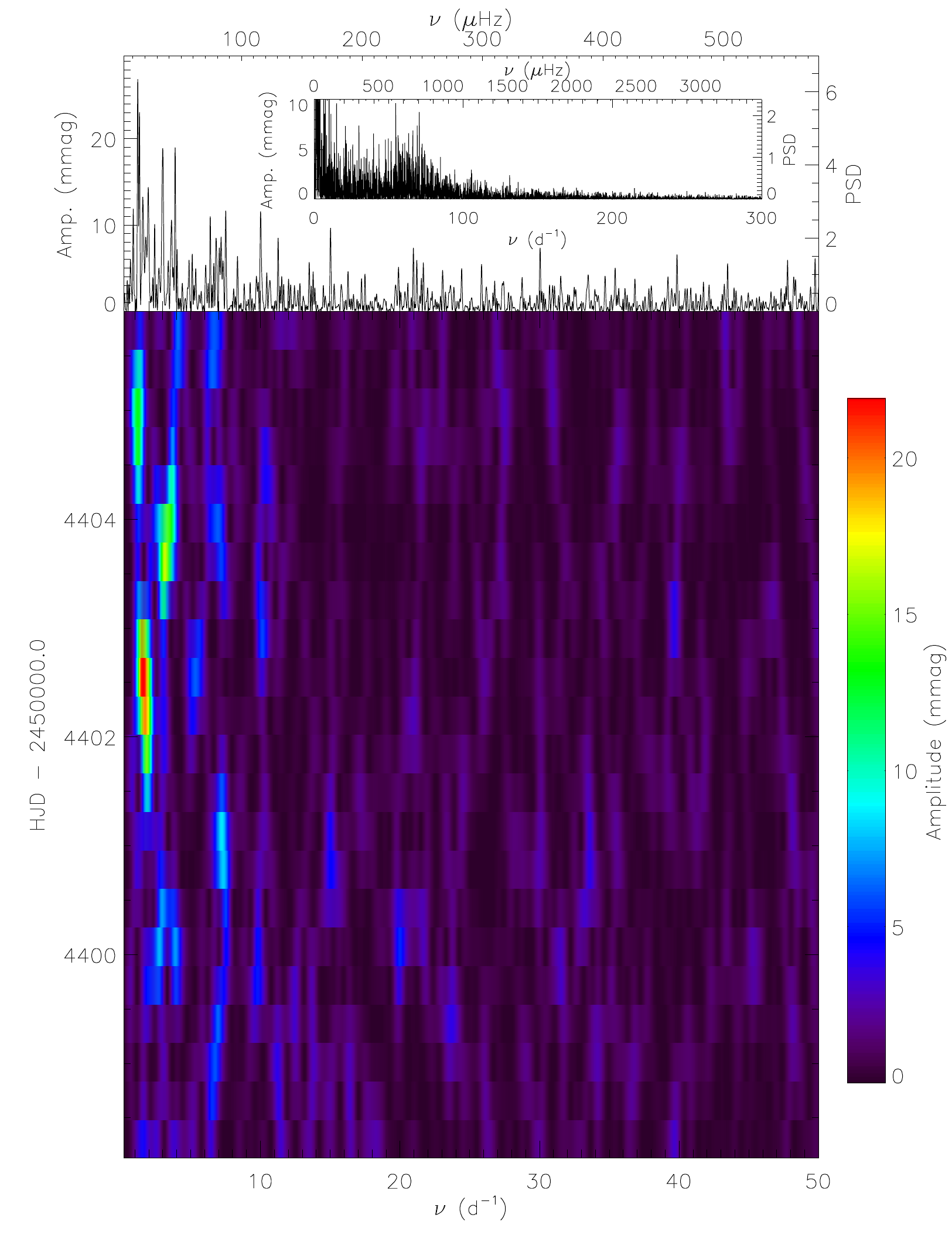}
\caption{{\it Upper panel}:~Fourier amplitude spectrum of the residuals from Figure\,\ref{resi}, with insert showing expanded scales. \new{The periodogram is marked in both amplitude (mmag) and PSD units.} {\it Lower panel}:~Time-frequency Fourier plot with 8-day running windows in time. The signal-to-noise of the highest peak (amp.\,=\,26.7 mmag, PSD\,=\,6.5 between HJD\,--\,2450000\,=\,4401--4403) is 3.36.}
\label{stft}
\end{figure}

In Figure \ref{resi} this mean superhump light curve is superimposed on the data (upper panel). The resulting residuals (lower panel) have been analyzed for other signals up to 300\,d$^{-1}$, or 288 Hz, (insert in the upper panel of Figure \ref{stft}), and up to 40\,d$^{-1}$, or 2160\,Hz, (Figure \ref{stft}, upper panel). \new{Both Fourier spectra show only noise with very low amplitude.} In order to search for time-dependent signals, we applied 8-day running windows in time in the Fourier amplitude spectra vs. time (Figure \ref{stft}, lower panel). There are three transient signals with marginal significance: \\

\indent\indent$f_1=1.204\pm0.002 \rm{d}^{-1}$, S/N\,$=3.36$\\
\indent\indent\indent   ($=13.93 \pm 0.02 \mu$Hz, P = 0.831 d)\\
\indent\indent$f_2=1.321\pm0.002 \rm{d}^{-1}$, S/N\,$=3.05$\\
\indent\indent\indent   ($=15.29 \pm 0.02 \mu$Hz, P = 0.757 d)\\
\indent\indent$f_3=3.884\pm0.002 \rm{d}^{-1}$, S/N\,$=2.81$)\\
\indent\indent\indent   ($=44.95 \pm 0.02 \mu$Hz, P = 0.257 d)\\

\new{Our frequency analysis did not reveal any evidence for quasi-periodic oscillations (QPOs) in the period range 15 -- 30 minutes (48 -- 96 d$^{-1}$, or 556 -- 1111 $\mu$Hz), with an upper limit of mean semi-amplitude of 0.25\%. Neither during the entire run, nor in limited time intervals, could we identify significant QPOs.  Kim et al. \cite{Ki09} found occasional QPOs with mean semi-amplitudes between 1.4 and 8.2\% in the period range between 10 and 27 minutes, which apparently were not present in our MOST observations. They only show random variability in typical time scales of a about 10--20 minutes with amplitudes up to 0.1\,mag (flickering). }

\section{Discussion}\label{Discussion}

Photometric variability of TT\,Ari with time resolution of a few minutes has been occasionally observed for more than 5 decades. A historical record of superhump periodicities was listed in Table 3 of Tremko et al. (\cite{Tr96}). According to this, TT\,Ari had negative superhumps between 1961 and 1978, with periods between 0.1326 and 0.1329 d. In May 1979 it entered into a low state phase which lasted 6 years in total, ending in May 1985. Between 1986 and 1996 several authors (Skillman et al. \cite{Sk98}; Andronov et al. \cite{An99}; Kraicheva et al. \cite{Kr99}) again observed negative superhumps with periods between 0.13276 and 0.13298 d. Afterwards, TT\,Ari switched to positive superhumps, with periods $\sim$0.149 d  before the end of 1997 and maintained them at least till  December 2004 (Skillman et al. \cite{Sk98}; Wu et al. \cite{Wu02}; Andronov et al. \cite{An05}). In October 2005 the negative superhumps had reappeared: Kim et al. (\cite{Ki09}) determined at the beginning of 2006 that the period, P$_{\rm{S}} = 0.132322 \pm 0.000053$\,d, was significantly smaller than any other period observed before. Only 19 months after Kim et al. (\cite{Ki09})'s observations were the MOST data presented here obtained, with P$_{\rm{S}} = 0.133010 \pm 0.000036$\,d, much larger than Kim et al's value. In July 2009, 20 months after the MOST observations, TT\,Ari entered into a new low state, returning to its normal brightness in January 2011. Afterwards and in the following 2012 season TT\,Ari was observed by many AAVSO members, including time-resolved CCD photometry. A preliminary analysis of these data reveals a rather stable superhump period P$_{\rm{S}} = 0.132883 \pm 0.000003$\,d (Vogt et al. in preparation), within the range of the historical observations between 1961 and 1988.  Therefore, the superhump period presented here for October/November 2007 is larger than any other value known before or after the MOST run. Perhaps there is some stronger variation in P$_{\rm{S}}$ during the last few years before TT\,Ari entered into a low state, while much more stability is achieved during the time after returning to its normal brightness. 

The current theoretical view on the superhump phenomenon can be summarized in the following way: Positive superhumps are probably due to development of an eccentric configuration in the accretion disk, but a mechanism for generating the superhump light is still under debate, for instance the periodic variation in the infall distance of the mass-transfer stream (Vogt \cite{Vo82}) or the periodic tidal stresses in the disk (Whitehurst \cite{Wh88}). The excitation of negative superhumps must have another mechanism, arising perhaps from a different type of precessional motion: the accretion disk wobble (Skillman et al. \cite{Sk98}).  If something can drive the disk out of the orbital plane, then the torque from the secondary will cause the disk to wobble backward; the nodes will regress, causing the geometry between disk and secondary to recur on a period slightly less than P$_{\rm{0}}$. There is a way to test this photometrically in TT\,Ari: a search for the expected nodal-precession signal at $\sim$4.2 days, the beat period between P$_{\rm{0}}$ and P$_{\rm{S}}$ of negative superhumps. However, the total duration of our MOST run of only 10 days ($\sim$2 beat periods)  may be too short for a significant detection of such a signal. Fortunately, we were successful in getting a new {\it MOST} light curve in 2012 that spans over 38 days of \new{nearly} contiguous observations. This new light curve will be presented and analyzed in a forthcoming paper.

\begin{figure}
\includegraphics[width=8cm]{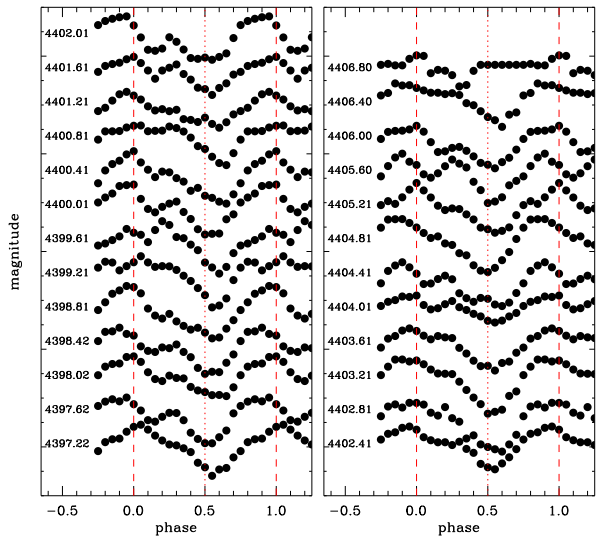}
\caption{Light curves of the entire MOST run vs. phase of ephemeris (\ref{supmax}). Each curve presents the average of 3 subsequent superhump cycles of TT\,Ari, binned in 0.05 phase intervals. The magnitude scale is marked in units of 0.1 mag, with arbitrary zero points. The numbers indicate  the first epoch of phase 0 in each light curve. There are strong variations in amplitude and light curve shape, even at these relatively short time scales of less than 10 hours between one light curve and the next one, underlining the importance of the uninterrupted observing mode of the MOST satellite. }
\label{changes}
\end{figure}

\begin{table}
 \centering
\caption{Superhump maxima and minima in the MOST light curves of TT Ari.}
\label{tabsup}
\begin{tabular}{cccccc}\hline
E & \multicolumn{2}{c}{HJD - 2454000} & E & \multicolumn{2}{c}{HJD - 2454000}\\ 
 & (Max.) & (Min.) & & (Max.) & (Min.) \\
\hline
0      &      397.217      &       397.151 & 37    &      402.151       &               -        \\
1      &      397.351      &       397.283 & 38    &      402.270       &       402.209\\
2      &      397.488      &       397.434 & 39    &      402.404       &               -        \\
3      &      397.624      &       397.556 & 40    &      402.538       &       402.480\\
4      &      397.752      &       397.693 & 41    &      402.656       &               -        \\
5      &      397.888      &       397.815 & 42    &           -                &       402.7560\\
6      &      398.006      &       397.964 & 43    &           -                &       402.900\\
7      &      398.146      &       398.114 & 45    &      403.185       &       403.157\\
8      &            -              &       398.239 & 46    &      403.327       &       403.280\\
10    &            -              &       398.495 & 47    &      403.476       &       403.414\\
11    &      398.692      &       398.636 & 48    &      403.602       &       403.551\\
12    &      398.816      &             -         & 49    &      403.767       &       403.678\\
13    &      398.939      &             -         & 50    &               -            &       403.811\\
14    &      399.068      &       399.036 & 52    &      404.099       &       404.092\\
15    &      399.224      &       399.162 & 53    &      404.293       &               -        \\
16    &       399.359     &       399.287 & 54    &      404.428       &       404.371\\
17    &       399.470     &       399.433 & 56    &      404.669       &       404.623\\
18    &           -               &       399.564 & 58    &      404.928       &       404.872\\
21    &      400.041      &              -        & 59    &      405.075       &       405.016\\
23    &      400.283      &              -        & 60    &      405.197       &       405.135\\
24    &      400.415      &       400.367 & 61    &      405.348       &              -        \\
25    &      400.539      &             -         & 62    &      405.484       &       405.426\\
26    &      400.702      &       400.641 & 63    &            -               &       405.550\\
28    &      400.956      &       400.902 & 64    &      405.726       &       405.690\\
29    &          -                &       401.031 & 65    &      405.872       &              -        \\
30    &      401.231      &              -        & 66    &      405.997       &       405.951\\
31    &          -                &       401.295 & 67    &      406.147       &       406.074\\
32    &      401.469      &               -       & 68    &      406.256       &       406.217\\
33    &          -                &       401.550 & 69    &      406.387       &              -        \\
34    &      401.746      &       401.680 & 70    &      406.555       &       406.482\\
35    &           -               &       401.816 & 71    &      406.663       &              -        \\
36    &      402.030      &       401.965 & 72    &      406.795       &       406.744\\
\hline
\end{tabular}
\end{table}

\acknowledgements

NV acknowledges the support by project Gemini-CONICYT 32090027 and DIUV 38/2011. ANC gratefully acknowledges support from the Chilean Centro de Astrof\'isica FONDAP No.15010003, the Chilean Centro de Excelencia en Astrof\'isica y Tecnolog\'ias Afines (CATA) BASAL PFB-06/2007, the Comite Mixto ESO-Gobierno de Chile and GEMINI-CONICYT No. 32110005. DBG, JMM, AFJM and SMR acknowledge financial support from NSERC (Canada) and for AFJM also FQRNT (Qu\'ebec). RK and WWW acknowledge support by the Austrian Science Fonds (FWF P22691-N16).

\newpage



\end{document}